# On the thermalization achieved in heavy-ion collisions

Sukhjit Kaur[1], Aman D. Sood[2] and Rajeev K. Puri[1*]

[1]*Physics Department, Panjab University Chandigarh -160014, INDIA*
[2]*SUBATECH,
Laboratoire de Physique Subatomique et des Technologies Associées
Université de Nantes - IN2P3/CNRS - EMN
4 rue Alfred Kastler, F-44072 Nantes, FRANCE*
*email: rkpuri@pu.ac.in

## Introduction

The heavy-ion collision at intermediate energies is excellent tool to study the nuclear matter at high density and temperature. At high excitation energies, the colliding nuclei not only compress each other, they also heat the matter [1]. As a result, the destruction of initial correlations takes place, which makes the matter homogeneous and one can have global stopping. The global stopping is defined as the randomization of one-body momentum space or memory loss of the incoming momentum. The degree of stopping, however, may vary drastically with incident energies, mass of the colliding nuclei and colliding geometry. The degree of global stopping has also been linked with the thermalization (equilibrium) in heavy-ion collisions. More the initial memory of the nucleons is lost, better it is stopped and themalized.

We here plan to investigate the degree of stopping reached in heavy-ion reactions.

## The Model

The present study is carried out within the framework of Quantum Molecular Dynamics (QMD) model [2,3]. The QMD model is an n-body theory which simulates the heavy-ion reactions at intermediate energies on event by event basis. The fragments are identified using modified Minimum Spanning Tree method where in addition to spatial correlations, fragments should also be bounded by the binding energy:

$$\zeta_i = \frac{1}{N^f}\sum_{i=1}^{N^f}\left[\frac{(\mathbf{p}_i - \mathbf{P}^{cm})^2}{2m_i} + \frac{1}{2}\sum_{j\neq i}^{N^f}V_{ij}(\mathbf{r}_i,\mathbf{r}_j)\right] < E_{bind} \quad (1)$$

We take $E_{bind}$ = - 4.0 MeV if $N^f \geq 3$ and $E_{bind}$ = 0.0 otherwise. Here $N^f$ is the number of nucleons in a fragment and $\mathbf{P}^{cm}$ is the center-of-mass momentum of a fragment. This method is known as Minimum Spanning Tree method with binding energy check (MSTB) [3].

## Results and Discussion

We performed the simulations for central reactions of $^{20}$Ne + $^{20}$Ne, $^{40}$Ar + $^{45}$Sc, $^{58}$Ni + $^{58}$Ni, $^{86}$Kr + $^{93}$Nb, $^{129}$Xe + $^{118}$Sn, $^{86}$Kr + $^{197}$Au and $^{197}$Au + $^{197}$Au using hard equation of state along with Cugnon cross section. The incident energies used in the above reactions are the energies at which the maximal production of intermediate mass fragments (IMFs) occurs.

We shall here consider few different quantities capable of estimating the degree of global stopping or thermalization. The relative momentum $<K_R>$ and anisotropy ratio $<R_a>$ are closely related to the degree of thermalization. The average relative momentum is defined as [4]:

$$\langle K_R \rangle = \langle |P_P(\mathbf{r},t) - P_T(\mathbf{r},t)|/\hbar \rangle, \quad (2)$$

where

$$P_k(\mathbf{r},t) = \frac{\sum_{j=1}^{A_k} P_j(t)\rho_j(\mathbf{r},t)}{\rho_k(\mathbf{r},t)}. \quad (3)$$

Here $P_j$ and $\rho_j$ are the momentum and density experienced by the $j^{th}$ particle and $k$ stands for either target or projectile. The relative momentum depends strongly on the local position $r$ and thus, is an indicator of local equili-brium. In fig. 1(a), we display the time evolution of the relative momentum $<K_R>$. Due to high density in heavier systems, $<K_R>$ remains finite for longer time. This shows that the equilibrium could be reached better in heavier systems compared to lighter ones.

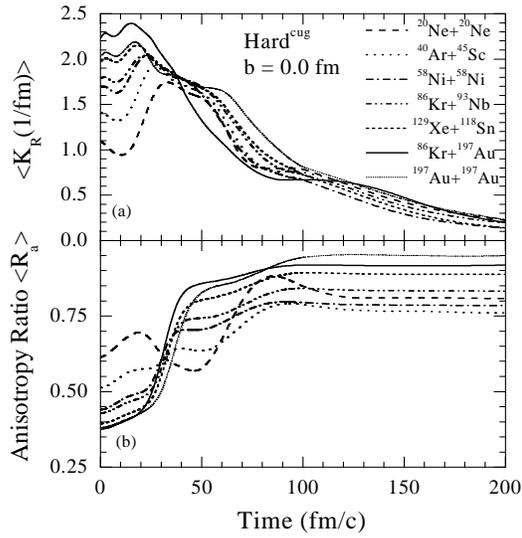
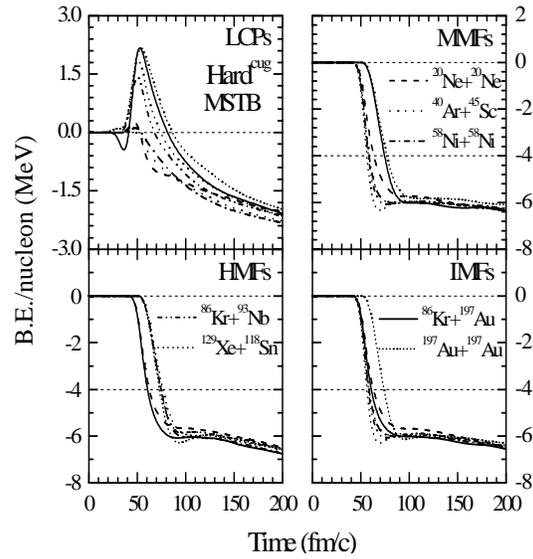

**Fig.** 1: The time evolution of (a) relative momentum and (b) anisotropy ratio. Here the results shown are for seven entrance channels at their corresponding peak energies.

The second quantity is anisotropy ratio which is defined as [4,5]:

$$\langle R_a \rangle = \frac{\sqrt{\langle p_x^2 \rangle} + \sqrt{\langle p_y^2 \rangle}}{2\sqrt{\langle p_z^2 \rangle}}. \qquad (4)$$

The anisotropy ratio $<R_a>$ is an indicator of global equilibrium of the system. This quantity does not depend upon the local density and thus represent the equilibrium of the whole system. The full equilibrium is achieved when $<R_a>$ is close to unity which could be achieved only through head-on collisions (b = 0.0 fm). In fig. 1(b), we display the time evolution of anisotropy ratio. It is clear from the fig. that equilibrium could be achieved in heavier systems compared to lighter ones. During the high density phase, anisotropy ratio changes drastically. Once the high density phase has passed, no more changes occur in thermalization.

**Fig.** 2: Same as Fig. 1, but for time evolution of binding energy per nucleon.

In fig. 2, we display the binding energy per nucleon of light charged particles (LCPs), medium mass fragments (MMFs), heavy mass fragments (HMFs) as well as IMFs for the reactions discussed above. We observed that at 200 fm/c, small fragments are still not cold and they take very long time to cool down. Whereas the heavy fragments are properly bound. The frozen time scale comes out to be same in both figures.

## Acknowledgments

This work is supported by Indo-French project.